\documentclass[12pt]{spieman}  % 12pt font required by SPIE;
\usepackage{amsmath,amsfonts,amssymb}
\usepackage{graphicx}
\usepackage{setspace}
\usepackage{tocloft}
\usepackage{lineno}
\usepackage{subcaption}
\usepackage{caption} 
\usepackage{multirow}
\usepackage{hyperref}% makes DOI clickable
\usepackage{comment}
\usepackage[dvipsnames]{xcolor}
\usepackage[inline]{enumitem}
\usepackage{tikz}
\usetikzlibrary{shapes.geometric, arrows, positioning}

\tikzstyle{startstop} = [rectangle, rounded corners, minimum width=3cm, minimum height=1cm, text centered, draw=black, fill=red!30]
\tikzstyle{process} = [rectangle, minimum width=3cm, minimum height=1cm, text centered, draw=black, fill=blue!20]
\tikzstyle{decision} = [diamond, minimum width=3cm, minimum height=1cm, text centered, draw=black, fill=green!20]
\tikzstyle{arrow} = [thick, ->, >=latex]

\title{Chang’e~7 Lunar Lander Optical Camera-Telescope: Optical Astronomy from the Moon}

\author[a]{Quentin Andrew Parker}
\author[a,d*]{Partha Sarathi Pal}
\author[5]{Junhao Chen}
\author[a]{Meng Su}
\author[a]{SeyedAbdolreza Sadjadi}
\author[a]{Andreas Ritter}
\author[a]{Andy~C.~T.~Kong}
\author[a]{Zhengjie Tian}
\author[a]{Haoyang Yuan}
\author[c]{Suijian Xue}
\author[b$\dag$]{Steve Durst}

\affil[a]{Laboratory for Space Research, Faculty of Science, The University of Hong Kong, Hong Kong SAR 99077, China}
\affil[b]{International Lunar Observatory Association, Kamuela, HI 96743, USA}
\affil[c]{National Astronomical Observatories, Chinese Academy of Sciences, Beijing 100012, China}
\affil[d]{Institute of Astronomy Space and Earth Science, Kolkata, India}
\affil[e]{Eastside Preparatory School, 10613 NE 38th Pl, Kirkland, WA 98033 USA}
\cftpagenumbersoff{figure}
\cftpagenumbersoff{table} 
\begin{document} 
\maketitle

%%%%%% Abstract %%%%%%
\begin{abstract}
We report the design and manufacture of a new, lightweight, wide-field, optical camera-telescope on 
board the Chang'e 7 lunar mission (launched in August 2026 and due for lunar touchdown in late November 2026). 
The camera is capable of static, 
panchromatic imagery within a $420-696~nm$ optical wavelength range. 
The camera was designed and built 
under the small lunar astronomy observation station program of the Chinese National Space Agency (CNSA) 
as a collaboration between the International Lunar Observatory Association  of Hawaii (ILOA), the 
Laboratory for Space Research (LSR) of the University of Hong Kong (HKU) and  the Beijing Institute of 
Space Mechanics and Electricity (BISME). The camera has been built to meet science goals of the 
mission for sustainable astronomical operation over a large range of temperatures from the Moon's south pole. We 
report on the design and ground based preliminary performance, together with an analysis of the 
camera's simulated output to indicate the range of astronomical observations possible from the lunar 
surface given the camera's limited sensitivity and angular resolution given the modest aperture and wide field of view. 
\end{abstract}

% Include a list of up to six keywords after the abstract

% Include email contact information for corresponding author
{\noindent \footnotesize\textbf{*}Partha Sarathi Pal,  \linkable{partha@hku.hk}}\\
{\noindent \footnotesize\textbf{$\dag$}Steve Durst unfortunately passed away in January, 2026. We include him in the author list, in order to pay respect to his involvement to the project till the last day of his life.}

%\begin{spacing}{2}   % use double spacing for rest of manuscript

\section{Introduction}
The Moon offers a unique and superior platform for astronomical observation and stable remote sensing 
of Earth, free from atmospheric turbulence, tropospheric extinction, and anthropogenic light 
pollution that limit ground-based astronomical facilities. Lunar-based astronomical observation 
offers unique astrophysical advantages, including ultra-long continuous integration enabled by the 
slow rotation of the Moon and the shielding of radio-frequency interference (RFI) possible on the 
lunar far side, permitting access to wavebands inaccessible to ground- and near-Earth space telescopes. 

Lunar astronomical observations have taken place via multiple in-situ missions since the 
1970s, providing technical and scientific heritage for subsequent lunar observatory development. The 
Apollo 16 mission deployed the first lunar-borne far-ultraviolet (FUV) spectroscopic imager in 1972, 
conducting pioneering FUV astronomical observations from the lunar surface. This payload realized 
direct detection of interstellar hydrogen distribution and captured Earth’s geocorona FUV radiation, 
successfully demonstrating the scientific potential of extraterrestrial atmosphere-free ultraviolet 
observation. Nevertheless, the mission was limited by short extravehicular operation duration, manual 
reliance, and lack of autonomous long-term tracking capability, resulting in discontinuous and short-
period astronomical data acquisition \cite{1972E&PSL..17...36T, 1977JGR....82..737M, 1975LPSC....6.1363R, 2021LPI....52.1483M}.

%The Chang’e-3 mission deployed the first fully autonomous lunar ultraviolet telescope (LUT) in 2013 on the lunar near side, marking the arrival of long-term robotic lunar astronomical observation. Equipped with a Ritchey-Chrétien optical system covering the 245–340 nm near-ultraviolet band, the telescope achieved two-axis autonomous sky tracking and completed large-area near-ultraviolet sky surveys, constructing a dedicated lunar-based ultraviolet star catalogue. It also realized continuous monitoring of variable stars, active galactic nuclei, and stellar flare activities, while synchronously detecting Earth’s plasmasphere radiation. However, the observation efficiency was constrained by lunar diurnal cycles: solar-dominated power supply only supports effective operation during lunar daytime, with mandatory hibernation throughout long lunar nights. In addition, fixed installation limits sky coverage, and persistent lunar thermal cycling and dust erosion gradually degrade optical and detector performance during long-term in-orbit service.

A series of conceptual schemes proposed by NASA and ESA, including large-aperture lunar optical 
telescopes and lunar low-frequency interferometer arrays, further defined the technical framework of 
future-generation lunar astronomical facilities, though these schemes were not implemented due to 
program adjustments and cost.
Including previous lunar mission heritage, lunar-based astronomy possesses distinct observational 
advantages in ultraviolet photometry and spectroscopy, ultra-long, time-domain astronomy, and with 
low-frequency radio observations. At the same time, extreme surface thermal environments, dust 
contamination issues, energy supply limitations, and relay communication constraints remain the core 
technical challenges that restrict the performance of lunar astronomical telescopes. Solving the 
above bottlenecks is the key to realizing high-precision, long-duration, and large-field lunar 
astronomical observation in the future \cite{1990asee.nasaQ....O, 1991SPIE.1494..119C,1995JBIS...48...93N, 1996AdSpR..18k..43F,2008P&SS...56..368G,2024SPIE13095E..1JR, 2025AAS...24546104C, 2025AAS...24546105R, 2025arXiv250302105C, 2025epsc.conf.1806C}.

As a non-governmental international institution dedicated to advancing lunar-based astronomy, the 
International Lunar Observatory Association (ILOA \linkable{https://iloa.org/})
is a U.S.-based nonprofit scientific organization 
founded in 2007. It proposed and implemented a phased lunar observatory program to exploit the 
scientific potential of long-term in-situ observation at the lunar South Pole. Distinct from 
governmental space agencies, ILOA commits to promoting global civilian space cooperation, verifying 
spaceborne optical observation technologies, and establishing sustainable lunar astronomical 
observation systems \cite{2008LPICo1415.2071D}.

ILOA’s lunar observation roadmap consists of progressive technological verification and flagship 
mission deployment, including the ILO-X precursor mission, the current Chang'e 7 payload dubbed "ILO-C", 
where the "C" stands for International Collaboration, and the 
long-term ILO-1 flagship lunar observatory. The ILO-X mission completed the first private-funded 
lunar optical imaging test at the lunar South Pole in 2024, validating the environmental adaptability 
of miniature optical systems under extreme lunar conditions. As a key international cooperative 
payload selected for China’s Chang’e-7 mission, the ILO-C mission represents a multi-party collaborative project 
led by ILOA and the University of Hong Kong via it's Laboratory for Space 
Research (LSR) as principal co-funder. The National Astronomical Research Institute of Thailand (NARIT) 
has provided in-kind support,
%, together with the informal support of scientists from the National Astronomical Observatory of Chine (NAOC) - {\bf check if NAOC is even an official partner}. 
designed for wide-field astronomical observation at the lunar South Pole, ILO-C serves as an lunar surface 
demonstration for the planned ILO-1 permanent observatory, due for deployment around 2028 at Malapert 
Mountain to realize long-duration continuous lunar astronomical surveys and Earth environmental 
monitoring \cite{2008LPICo1446...50D, 2022AAS...24043104D}.

China has also emerged as one of the major space faring nations over the past two decades, achieving a 
series of remarkable milestones in space exploration. These include the historic Chang'e~ series of 
lunar missions with 1-6 already completed and Chang'e 7 and 8 in train 
\cite{2024NSRev..11D.329W}, the Tianwen-1 Mars mission \cite{2021AdSpR..67..812Z, 
2022JGRE..12707137W}, the indigenous Tiangong Space Station \cite{2023SpScT...3...35W}, 
and the world's largest radio telescope called the  Five-hundred-meter Aperture Spherical Telescope 
(FAST) \cite{2011IJMPD..20..989N}. These missions not only can be characterized as significant 
breakthroughs in the fields of space science and technology (reaching the far side of the Moon and 
uniquely returning 17~kg of moon rock) but also as a preparatory path for outer planetary exploration 
for the next generation (including plans to run a live tutorial class for students from the space 
station; See \url{https://sc.mp/g2zx?utm_source=copy-link&utm_campaign=3159118&utm_medium=share_widget}).   

%The China National Space Administration (CNSA) successfully completed a few missions with  along with a few upcoming space missions.
%The list of space missions consists of the construction of Tiangong, the Chinese Space Station, and other missions on Mars and Moon exploration.

\subsection{Chang'e Lunar Exploration Missions}
The Chang'e lunar exploration program, named after the mythological Chinese Moon goddess, represents 
one of the most ambitious and successful space exploration endeavors of the 21st century. The program 
started with Chang'e~1 in 2007. Chang'e~1 served as China's first lunar orbiter, mapping the Moon's 
surface and analyzing its geological composition \cite{ouyang2010}. Chang'e-2 was launched in 
2010, serving as an advanced orbiter that provided higher-resolution imagery and later extended its 
mission to explore the asteroid Toutatis \cite{2013NatSR...3.3411H}. Then in 2013, the program 
achieved a landmark milestone with Chang'e~3, which successfully deployed the Yutu rover, making 
China only the third nation to achieve a soft lunar landing \cite{ip2014}.
The Chang’e~3 mission deployed the first fully autonomous lunar ultraviolet telescope (LUT) in 2013 
on the lunar near side, marking the arrival of long-term robotic lunar astronomical observation. 
Equipped with a Ritchey-Chrétien optical system covering the 245–340 nm near-ultraviolet band, the 
telescope achieved two-axis autonomous sky tracking and completed large-area near-ultraviolet sky 
surveys, constructing a dedicated lunar-based ultraviolet star catalogue. It also realized continuous 
monitoring of variable stars, active galactic nuclei, and stellar flare activities, while 
synchronously detecting Earth’s plasmasphere radiation. However, the observation efficiency was 
constrained by the lunar diurnal cycle as the solar-dominated power supply can only supports 
effective operation during lunar daytime, with mandatory hibernation throughout the long lunar 
nights. In addition, fixed installation limits sky coverage, and persistent lunar thermal cycling and 
dust erosion gradually degraded the optical and detector performance during long-term 
in-situ service \cite{2016AdAst2016E...7Z, 2016Ap&SS.361...76Y}.

The next mission, Chang'e~4, launched in 2018, made history as the first spacecraft to land on the far 
side of the Moon in the Von Karman Crater, deploying the Yutu-2 rover and relying on the Queqiao 
relay satellite for moon-earth communication due to the lack of direct line of sight 
\cite{wu2017}. Chang'e~5 was launched in 2020, 
and it accomplished the first non-human lunar sample return mission since the Soviet Luna 24 mission 
in 1976, retrieving approximately 1.73 kg of lunar soil and providing new insights into the Moon's 
volcanic history \cite{li2022}. The Earth \& Planetary Science Department and the LSR of The 
University of Hong Kong are also actively involved in the analysis of the lunar Moon rock retrieved 
from the Chang'e~5 mission \cite{qian2021a, qian2021b}. 
Chang'e~6, launched in 2024, was China's ambitious mission to collect samples from the far side of 
the Moon, specifically targeting the South Pole-Aitken (SPA) Basin, one of the largest and oldest 
impact craters in the solar system. It relied on the Queqiao-2 relay satellite for communications 
\cite{2024AJ....168..247W}. The mission successfully returned approximately 1.93 kg of lunar far-side 
samples to Earth in June 2024, marking the first-ever sample return from the lunar far side and 
providing unprecedented insights into the geological history and compositional differences between 
the near and far sides of the Moon \cite{2024NSRev..11E.328L}.

Chang'e~7 is a comprehensive lunar exploration mission of the China National Space Administration 
(CNSA) as part of China's fourth phase of lunar exploration. Launched on August 24th 2026 on a 
Long~March~5 rocket from the Wenchang Satellite Launch Center on Hainan Island, China, the mission is 
designed to first conduct an in-orbit detailed survey of the lunar south pole. This is a region of 
significant scientific interest due to the potential presence of water ice in permanently shadowed 
craters, which could serve as a critical resource for future human lunar exploration 
\cite{2022Natur.610..308L}. The Chang'e~7 mission consists of five components: an orbiter, a 
lander, a rover, a mini flying probe, and another relay satellite, making it  the most complex non-
human robotic lunar missions ever attempted. The mini flying probe is particularly innovative, as it 
is designed to hop into permanently shadowed regions of the lunar south pole to directly detect the 
presence of water ice and other volatile substances \cite{2024NSRev..11D.329W, 
2020LPI....51.1755Z}. Chang'e~7 represents a significant step forward in China's long-term strategic 
goals for deep space exploration and the utilization of lunar resources \cite{ce7}. The proposed 
landing site for the lunar lander is at the peak near the southeast ridge of the Shackleton crater 
\cite{2011Icar..211.1066M,2023ScChD..66..417Z} and so where our ILO-C camera will be deployed.
The lunar lander consists of 7 different payloads dedicated to different experiments. 
The ILOA/HKU-LSR optical telescope is the only instrument on the lunar lander dedicated to exploring 
the potential of optical astronomy from the Moon.

\subsection{The ILOA-CNSA Long-Term Collaboration}
ILOA has maintained stable, mutually beneficial and in-depth scientific cooperation with China’s 
aerospace and astronomical communities for over a decade. Originating from a 2012 cooperation memorandum 
with Chinese astronomical institutions. The bilateral partnership achieved landmark lunar astronomical 
observation results based on the Chang’e-3 mission. The ongoing ILO-C Chang’e~7 cooperation further 
establishes a unique model of positive, non-governmental Sino-U.S. lunar scientific collaboration, 
avoiding governmental institutional constraints while advancing technological verification and 
scientific data sharing for lunar-based, global astronomy. Under the framework of China’s 
International Lunar Research Station, both parties intend to expand long-term cooperation in optical 
system calibration, observation algorithm optimization, and deep-space astronomical research, laying 
a foundation for the future construction and operation of international lunar observation facilities.

The ILO-C mission collaboration between ILOA and HKU-LSR grew from ILOA’s years of cooperation with 
Chinese space and astronomy communities dating back to the Chang’e~3 programme. When China issued an 
international payload call for Chang’e~7, ILOA took charge of the overall ILO-C proposal and 
international coordination, whereas HKU-LSR became the core science lead responsible for establishing 
camera technical specifications and the astronomical observations that are then possible. 
The Beijing Institute of Space Mechanics and Electricity 
(BISME; \linkable{https://www.cast.cn/english/channel/1808}), a sister organization under Chinese Academy of Space Technology \cite{bisme},  was 
chosen as the company to meet the technical camera hardware and optical specifications we established, 
including engineering implementation. The joint proposal was formally selected in 2024 within the Chang’e‑7 
hosted payload roster. Working alongside other partners, including  NARIT, ILOA and HKU-LSR have 
iterated the wide-field astronomical camera system for optimum operations at the lunar South Pole 
within the performance constraints of the achievable camera design. Besides 
pursuing key science targets such as observations of the Galactic Centre, ILO-C functions as a lunar 
technology demonstrator for ILOA’s planned ILO‑1 permanent lunar observatory, embodying a model of 
non-governmental international partnership in lunar exploration \cite{2026ChJSS..46..540X}.

\section{The ILO-C Wide Field Optical Telescope}
\begin{figure}[ht]
\centering
\begin{subfigure}{0.35\textwidth}
\centering
\includegraphics[width=\textwidth]{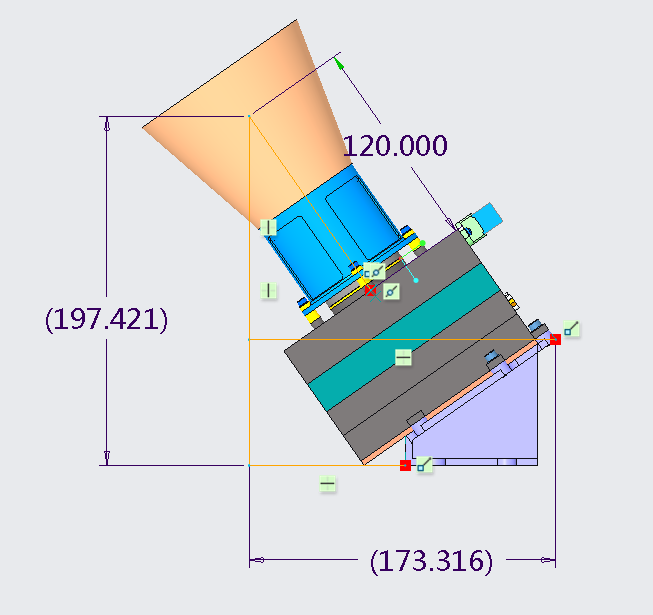}
\caption{}
\label{1a}
\end{subfigure}
\begin{subfigure}{0.52\textwidth}
\centering
\includegraphics[width=\textwidth]{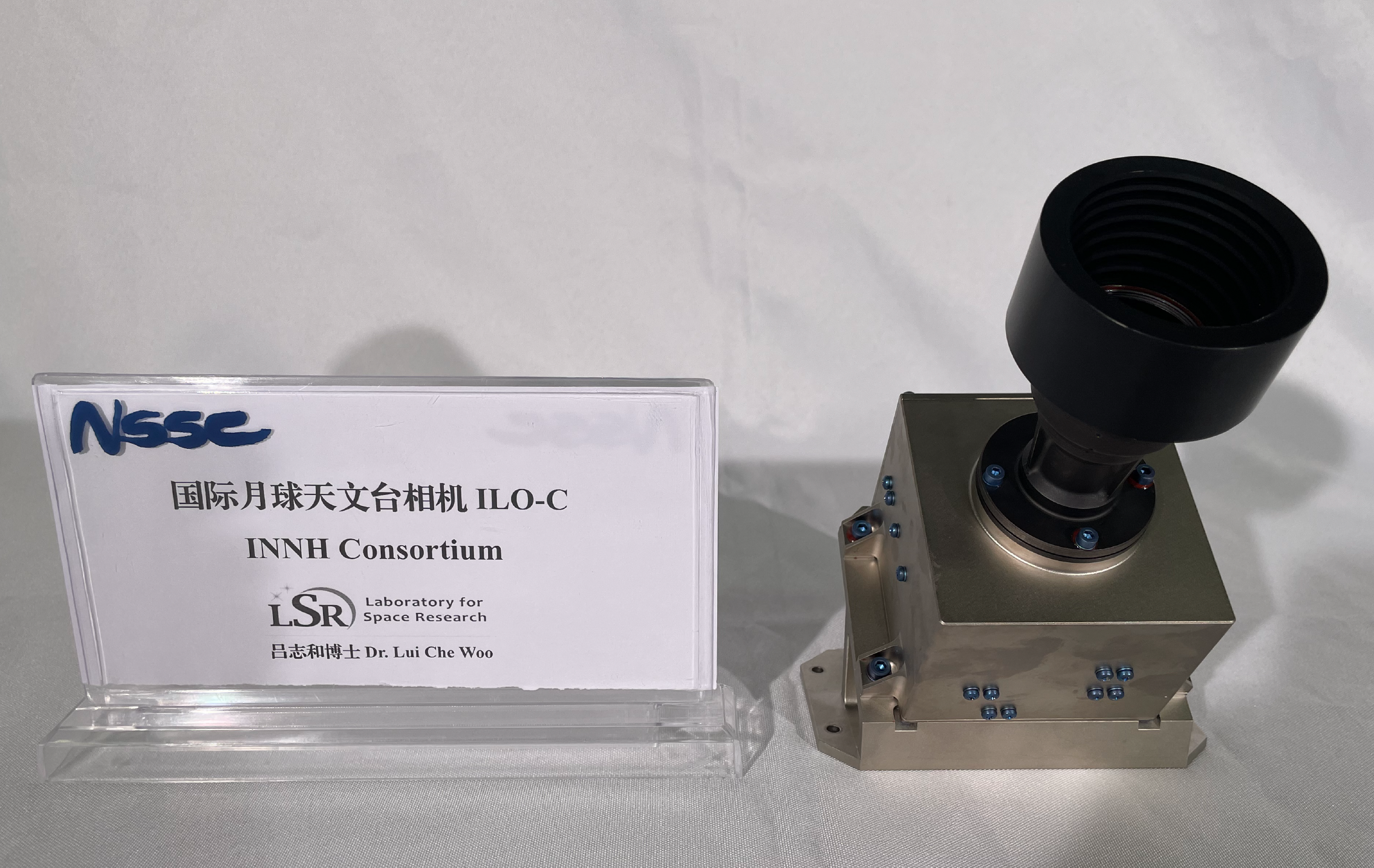}
\caption{}
\label{1b}
\end{subfigure}
\begin{subfigure}{0.55\textwidth}
\centering
\includegraphics[width=\textwidth]{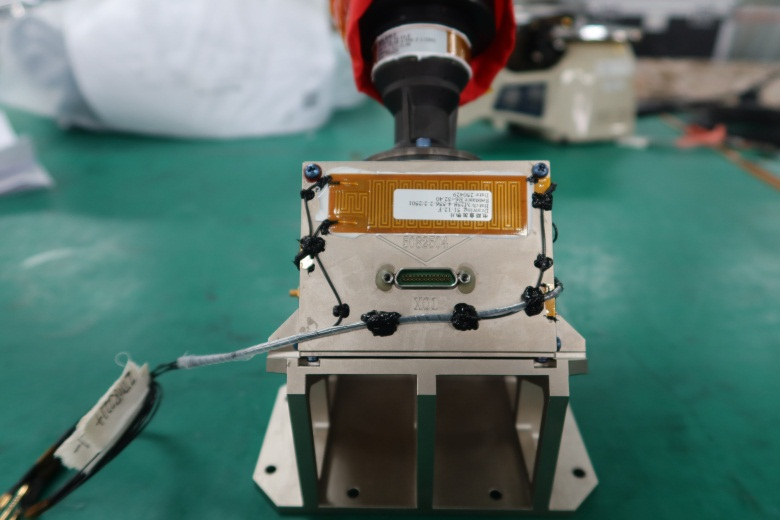}
\caption{}
\label{1c}
\end{subfigure}
\begin{subfigure}{0.32\textwidth}
\centering
\includegraphics[width=\textwidth]{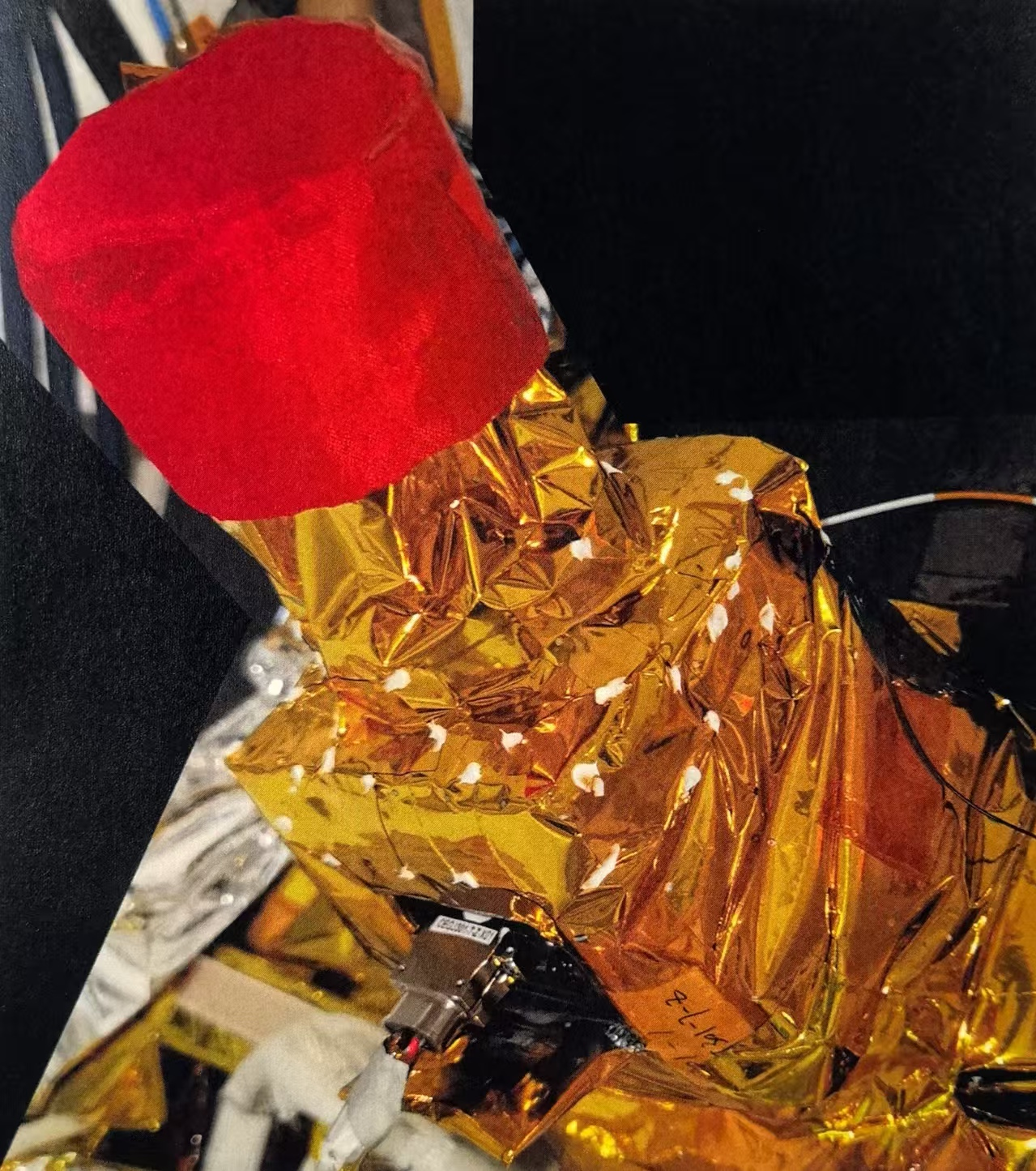}
\caption{}
\label{1d}
\end{subfigure}
\caption{HKU-LSR-ILOA camera at different stages of development. a) Schematic diagram and dimensions of the camera. b) Completed camera at calibration stage, See, Figure~3 of Ref.~\citenum{2026ChJSS..46..540X}. c) base of the camera showing the data readout system to the Lunar Lander. d) Final camera assembly covered with standard gold sheet thermal protection.}
\label{diagram}
\end{figure}

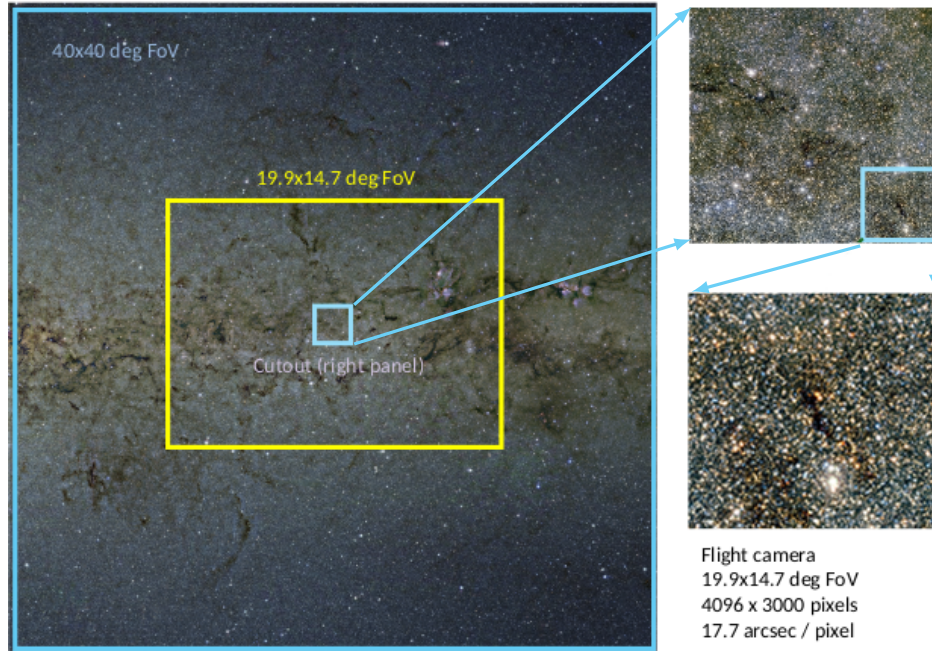
\begin{figure}[ht]
\centering
\begin{tikzpicture}
\node[anchor=south west, inner sep=0] (image) at (0,0) {
\includegraphics[width=0.75\textwidth]{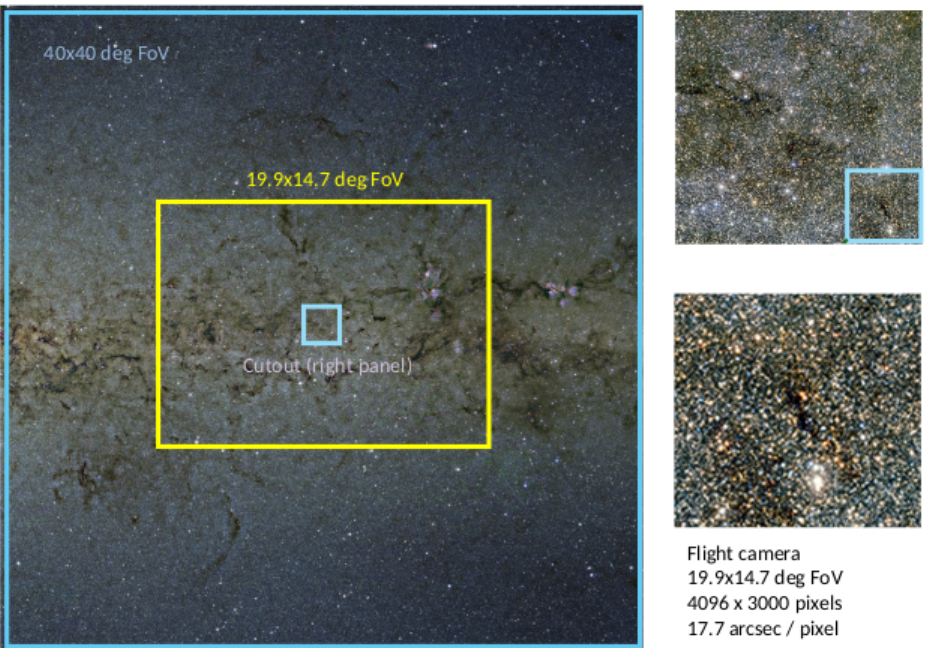}};
\begin{scope}[x={(image.south east)}, y={(image.north west)}]
\draw[->, very thick, >=latex, Cyan!50] (0.37, 0.53) -- (0.73, 0.985);
\draw[->, very thick, >=latex, Cyan!50] (0.37, 0.47) -- (0.73, 0.63);
\draw[->, very thick, >=latex, Cyan!50] (0.91, 0.62) -- (0.73, 0.55);
\draw[->, very thick, >=latex, Cyan!50] (0.99, 0.62) -- (0.99, 0.55);
\end{scope}
\end{tikzpicture}
\caption{Expected camera resolution and field of view in the context of 
anticipated ILO-C imaging of the Galactic plane. Base Image Credit:
ESO/VVV Survey/D. Minniti.}
\label{cam_res}
\end{figure}

%\section{The Overall Hardware}
The ILOA proposal to CNSA for astronomical observations from the lunar surface was made in response 
to an international call for science payloads on board Chang'E 7. The experimental concept was a small 
wide field optical astronomical telescope and it was subsequently selected by the CNSA of China in open 
competition along with other selected experiments. Due to some funding and technical difficulties, the LSR of HKU(\linkable{https://www.lsr.hku.hk/}) came on board shortly after to offer in-house scientific 
and technical expertise and to co-fund the mission. The optical telescope was designed and manufactured 
by BISME. 
%Beijing Institute of Space Mechanics and Electricity (BISME; \linkable{\url{https://www.cast.cn/english/channel/1808}), a sister organization under China Academy of Space Technology \cite{bisme}. 
Figure~\ref{diagram} shows the schematic diagram of 
the optical telescope and the photograph of the telescope at different stages of manufacturing. Stringent 
requirements for weight, overall size, and footprint for ILO-C had to be adhered to in order to match stringent CNSA payload mission profile requirements. Figure~\ref{cam_res} shows an anticipated field of fiew (yellow rectangle) of the ILO-C imaging the Galactic centre from the lunar surface.

%\subsection{CMOS Sensors in optical astronomy}
\subsection{The Detector}
The latest Complementary Metal-Oxide-Semiconductor (CMOS) detector is used in the ILO-C 
optical telescope electronics system. The CMOS detector technology has significantly transformed optical 
astronomy, offering compelling advantages over the traditionally dominant Charge-Coupled Device (CCD) 
sensors \cite{2024SPIE13103E..0RK}. Modern scientific CMOS (sCMOS) detectors provide superior 
readout speeds, lower read noise, higher dynamic range, and improved quantum efficiency across a broad 
wavelength range, making them particularly well-suited for time-domain astronomy, 
transient detection, and high-cadence observational surveys \cite{MAGNAN2003199, 2002ExA....14...33J, 
2013RAA....13..615Q}. The parallel readout architecture of CMOS sensors enables simultaneous multi-row 
processing, dramatically reducing readout times compared to CCD counterparts, which is critical for capturing 
fast astrophysical phenomena such as stellar occultations, exoplanet transits, and optical counterparts of 
gravitational wave events \cite{2020PASP..132l5001Z, 2022MNRAS.511.2405S, 2023PASP..135h5002B, 2026RASTI...5ag022A}. 
The on-flight and ground-based technical performance of sCMOS detectors is now being tested to understand the extent 
of technical improvement of sCMOS detector technology in order to achieve engineering excellence in astronomical 
observation projects \cite{2026AJ....171..283F, 2026PASP..138d5002W, 2025RMxAC..59....9S, 2025SPIE13527E..0VO}.

Given the stringent compact design needs for the ILO-C camera, as well as flight-verification experience, 
the image sensor uses an area-array SONY-IMX253 (\url{https://www.sony-semicon.com/files/62/flyer_industry/IMX253_255LLR_LQR_Flyer.pdf}) sCMOS chip. Recently, sCMOS sensor chips have 
been successfully applied in multiple on-orbit models, with extensive on-orbit flight experience giving us confidence
they are capable of meeting our mission requirements. The main technical specifications of the sCMOS detector chosen
are presented in the Table~\ref{cmos}.

\begin{table}[ht]
\caption{Specifications of the Image Sensor}
\label{cmos}
\begin{center}
%\scriptsize 
\begin{tabular}{|c|c|c|}
\hline\hline
\textbf{No} & \textbf{Parameter} & \textbf{Technical Specification} \\\hline
1 & Pixel Size & $3.45\,\mu\text{m} \times 3.45\,\mu\text{m}$ \\
\multirow{2}{*}{2} & \multirow{2}{*}{Resolution} & $4096(\text{H}) \times 3000(\text{V})$ \\
 & & $2048(\text{H}) \times 1500(\text{V})$ ($2\times2$ binning) \\
3 & Quantization Bits & 12-bit / 10-bit / 8-bit \\
4 & Readout Frame Rate & 64.6\,fps @ 10-bit \\
5 & Spectral Range & $420$--$700\,$nm \\
6 & Responsivity & $8.3\,$V/lux$\cdot$s \\
7 & Saturation Electron Count & $10359\,e^-$ (no binning) \\
8 & Dynamic Range & $>66\,$dB \\
9 & Noise (readout + dark, 200\,ms) & $2.4\,e^-$ rms \\
10 & Voltage Type & 3.3\,V, 1.8\,V, 1.2\,V \\
11 & Package & LGA \\
12 & Operating Temperature & $-30^{\circ}$C to $+75^{\circ}$C \\
13 & Storage Temperature & $-40^{\circ}$C to $+85^{\circ}$C \\
\hline
\end{tabular}
\end{center}
\end{table}

The ILO-C optical telescope is capable of static panchromatic imaging over the $420-696~nm$ wavelength range. 
The field of view (FoV) is $19^\circ \times 14^\circ$ as provided by the exceptional, well-designed, multi-element, lens 
array and was chosen as an effective compromise given the desire for wide field coverage, especially for the Galactic 
Plane, and the resulting, achievable angular resolution of $17.68''$. Each image will be captured using 
a $4096 \times 3000$ Bayer RGGB filter sCMOS sensor (\linkable{https://en.wikipedia.org/wiki/Bayer\_filter}). 
The expected sensitivity of the camera is $<10$ Magnitude with an expected SNR of 6. 

%\section{Optical telescope Data structure and data processing}
%\section{Performance Tests and Data Format}

\begin{figure}[ht]
\centering
\includegraphics[width=0.45\textwidth]{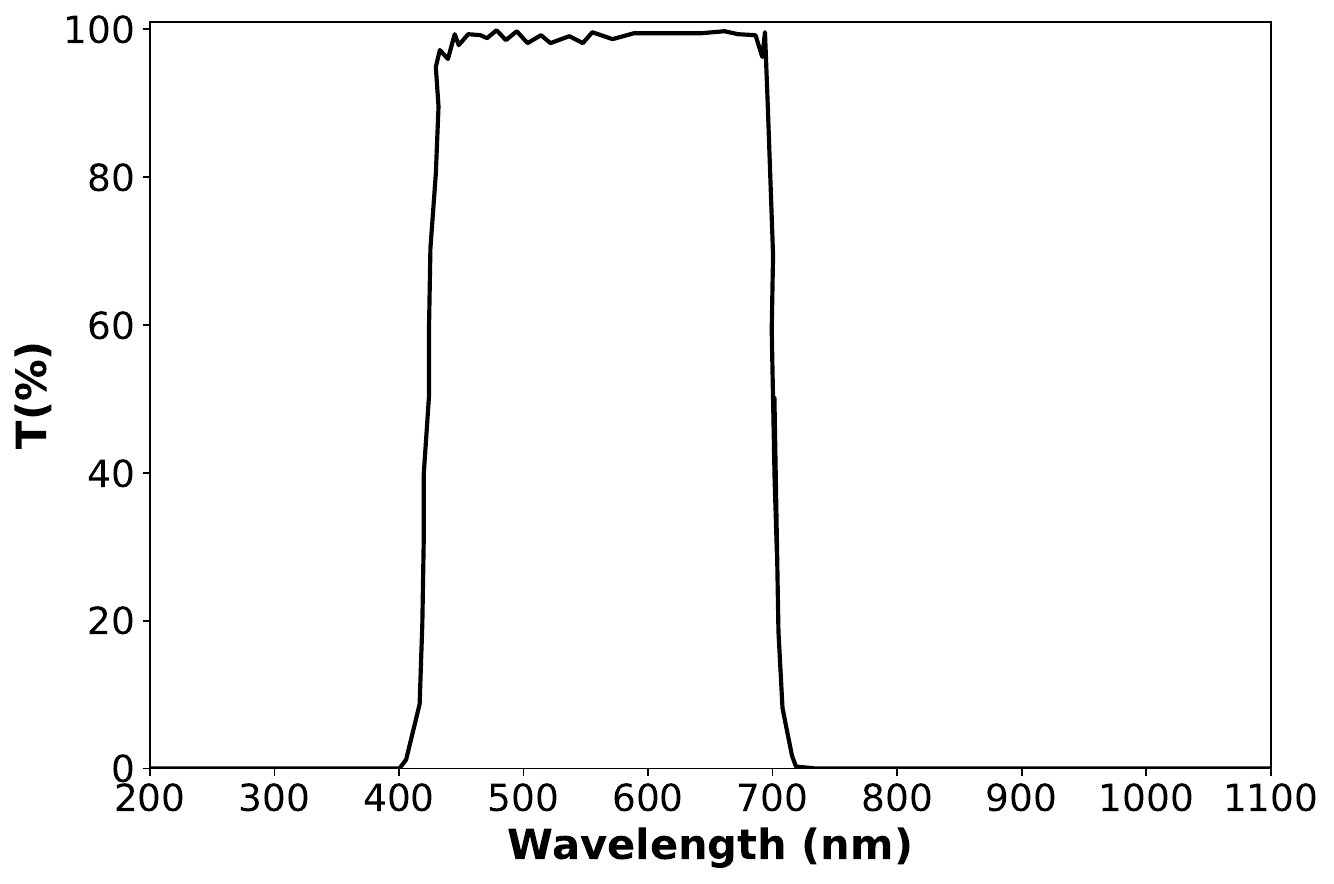}
\includegraphics[width=0.45\textwidth]{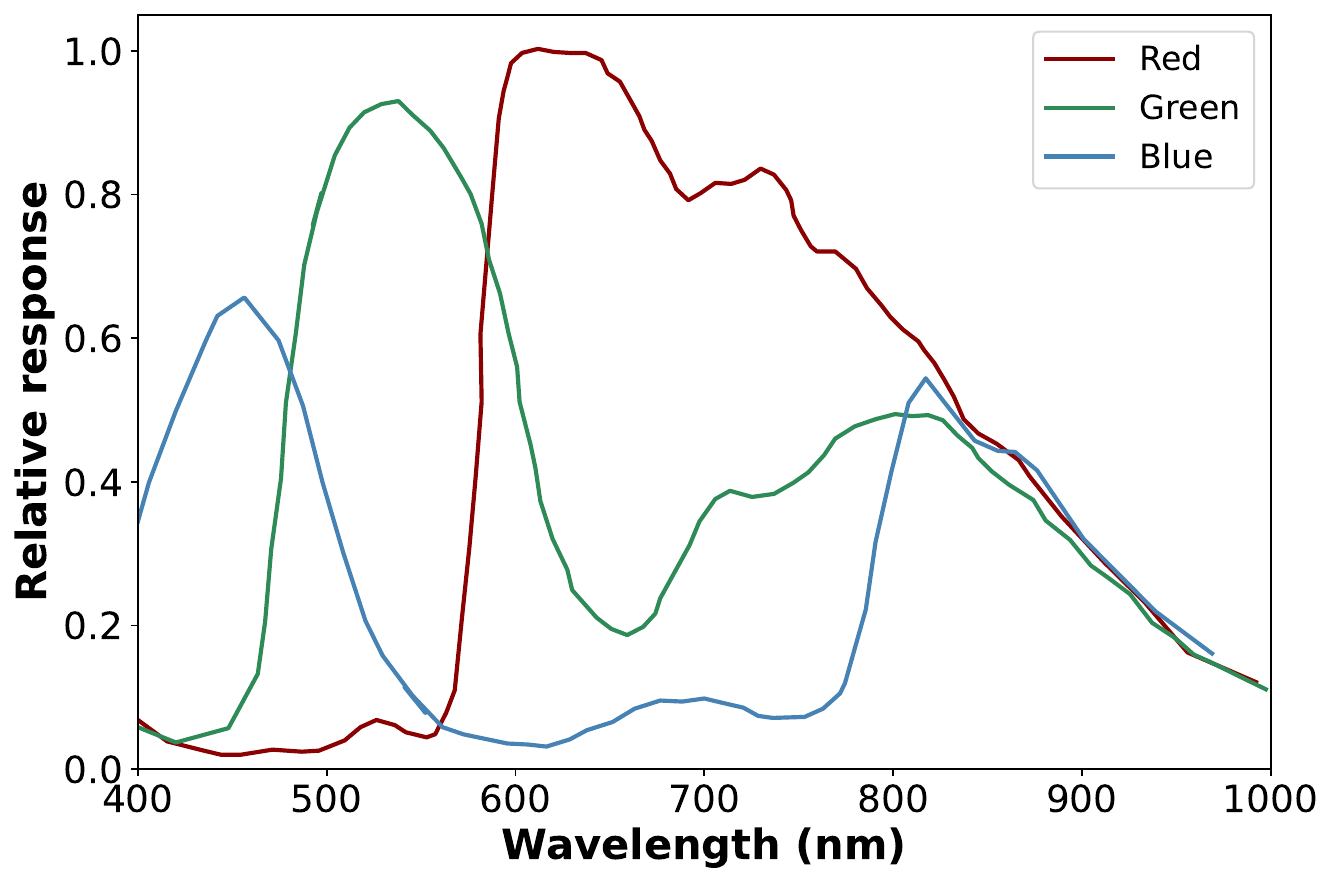}
\caption{The left panel shows the overall response function of the camera within the optical wavelength range. 
The right panel shows the camera response profiles for the RGB optical filters separately.}
\label{resp}
\end{figure}

Figure~\ref{resp} shows the overall spectral response curve of the instrument along with the optical 
Bayer RGB filter response functions of the sCMOS detector. The camera is capable of storing data at a rate of 50 
frames-per-second and $4096 \times 3000 \times 12$-bit storage per frame. The details of the camera laboratory 
determined performance are presented in Table~\ref{performance}.  

\begin{table}[ht]
\caption{ILO-C Camera Performance Indicators} 
\label{performance}
\begin{center}
\scriptsize 
\begin{tabular}{|c|c|c|c|c|}
\hline\hline
\textbf{No.} & \textbf{Performance Indicators} & \textbf{Required value} & \textbf{Measured value} & \textbf{Compliance} \\\hline
1 & Spectral range (nm) & 420--700 ($\pm$ 10) & 420--696 & Satisfies \\
2 & Imaging Mode & Static photography & Static imagery & Satisfies \\
3 & Field of view & $\geq$ 19$^\circ$ $\times$ 14$^\circ$ & 19.93$^\circ$ $\times$ 14.73$^\circ$ & Satisfies \\
4 & Number of effective pixels & $\geq$ 4096 $\times$ 3000 & 4096 $\times$ 3000 & Satisfies \\
5 & Angular resolution & Better than 20$''$ & 17.68$''$ & Satisfies \\
6 & Sensitivity (magnitude) & Better than 10th visual magnitude & 10 (By analysis) & Satisfies \\
7 & Signal-to-Noise Ratio & $\geq$ 3 & 5.96 (By analysis) & Satisfies \\
8 & Exposure time & Single exposure time $\leq$ 4s & 4s & Satisfies \\
9 & Image storage capability & $\geq$ 25 fps & 50 & Satisfies \\
10 & Quantization bit number & $\geq$ 10 bits & 12 & Satisfies \\
11 & Imaging power consumption & $\leq$ 10W (without thermal control) & 7.8 & Satisfies \\
12 & Weight & $\leq$ 1.4kg (tentative) & 1.18 & Satisfies \\
\hline
\end{tabular}
\end{center}
\end{table}

\begin{figure}[ht]
\centering
\includegraphics[width=0.75\textwidth]{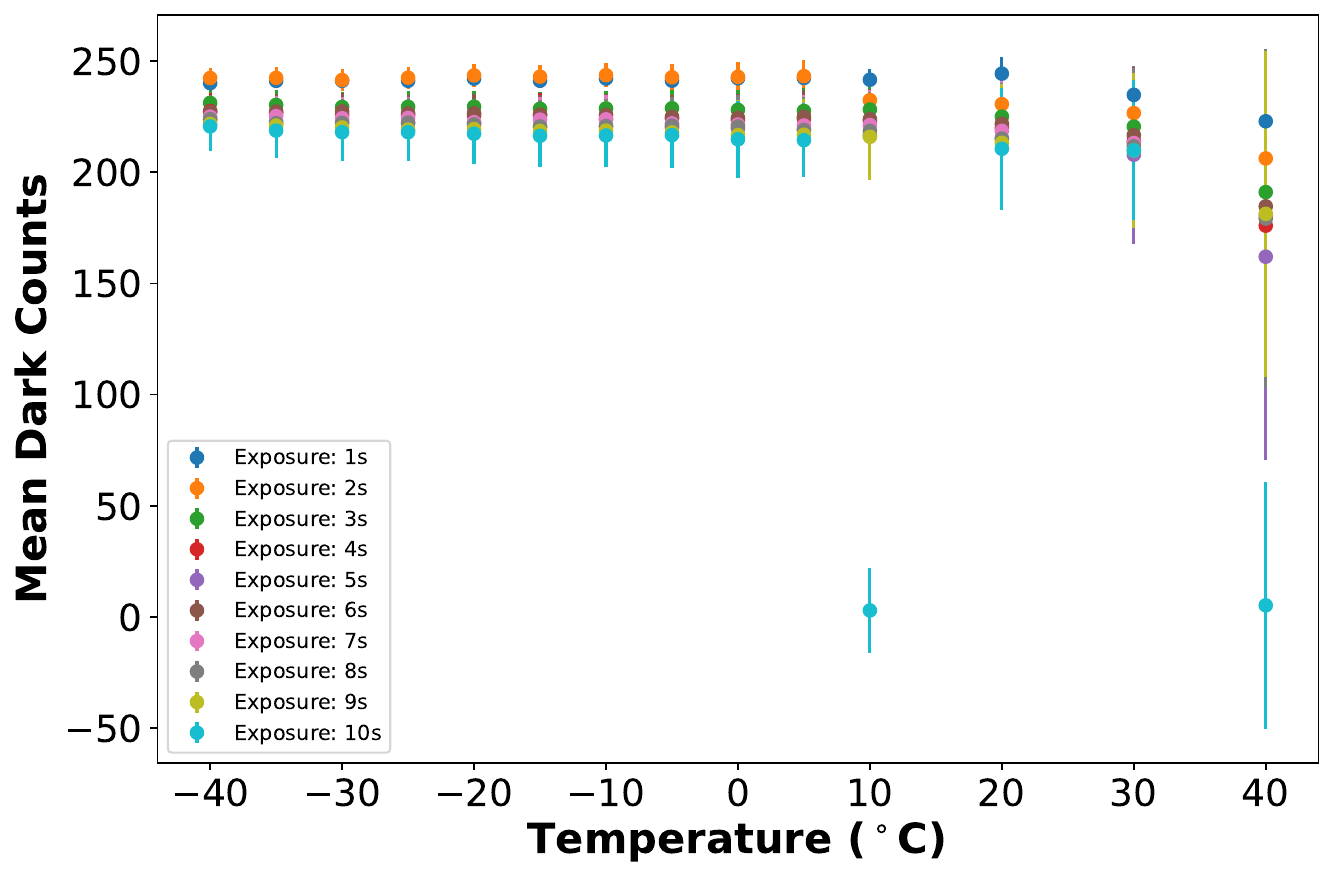}
\caption{Measured dark current at different exposure times and temperatures.}
\label{dark}
\end{figure}

BISME lab tested the sensor performance against the harsh temperature conditions close to those expected on the 
rim of the Shackelton Crater, the anticipated Chang'e~7 landing site. They reported an optimum working 
temperature range of -40$^{\circ}C$ to +40$^{\circ}C$. Camera dark current tests were performed at 5$^{\circ}C$ 
increments within this temperature range. At each temperature step, tests were conducted with exposure times 
of 10s, 9s, 8s, 7s, 6s, 5s, 4s, 3s, 2s, and 1s, and five image frames were saved for each exposure time. 
The instrument data obtained due to the sensor's dark current are consistently confined to the range 200-250 
counts over the temperature range from -40$^{\circ}C$ to +10$^{\circ}C$. However, at temperatures 
above +10$^{\circ}C$, a wider range of fluctuations in the dark current readings is observed across all 
exposure times. These results are shown in Fig.~\ref {dark}. 

\section{The Image Frame Processing Pipeline}

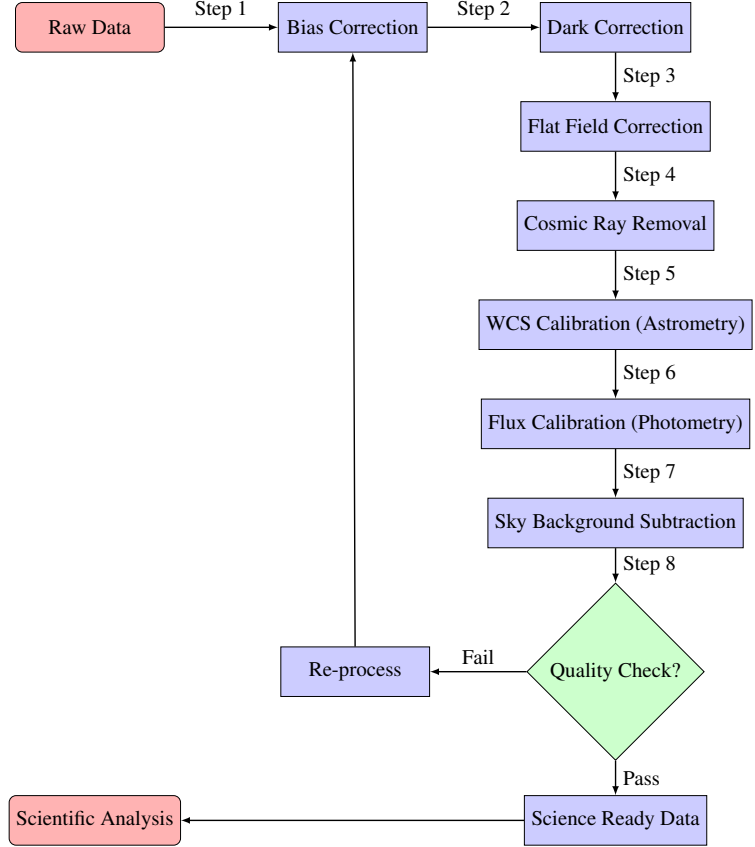
\begin{figure}[h]
\centering
\resizebox{0.6\textwidth}{!}{
\begin{tikzpicture}[node distance=1.5cm]
% Nodes
\node (rawdata) [startstop] {Raw Data};
\node (bias) [process, right of=rawdata, xshift=3.8cm, align=center] {Bias Correction};
\node (dark) [process, right of=bias, xshift=3.8cm, align=center] {Dark Correction};
\node (flat) [process, below of=dark, yshift=-0.5cm, align=center] {Flat Field Correction};
\node (cosmic) [process, below of=flat, yshift=-0.5cm, align=center] {Cosmic Ray Removal};
\node (wcs) [process, below of=cosmic,  yshift=-0.5cm, align=center] {WCS Calibration (Astrometry)};
\node (flux) [process, below of=wcs,  yshift=-0.5cm, align=center] {Flux Calibration (Photometry)};
\node (sky) [process, below of=flux,  yshift=-0.5cm, align=center] {Sky Background Subtraction};
\node (check) [decision, below of=sky, yshift=-1.5cm, align=center] {Quality  Check?};
\node (reprocess) [process, left of=check, xshift=-3.75cm, align=center] {Re-process};
\node (science) [process, below of=check, yshift=-1.5cm, align=center] {Science Ready Data};
\node (analysis) [startstop, left of=science, xshift=-9cm, align=center] {Scientific Analysis};

% Arrows
\draw [arrow] (rawdata) -- node[pos=0.5, above] {Step 1} (bias);
\draw [arrow] (bias)    -- node[pos=0.5, above] {Step 2} (dark);
\draw [arrow] (dark)    -- node[right] {Step 3} (flat);
\draw [arrow] (flat)    -- node[right] {Step 4} (cosmic);
\draw [arrow] (cosmic)  -- node[right] {Step 5} (wcs);
\draw [arrow] (wcs)     -- node[right] {Step 6} (flux);
\draw [arrow] (flux)    -- node[right] {Step 7} (sky);
\draw [arrow] (sky)     -- node[right] {Step 8} (check);
%\draw [arrow] (rawdata) -- (analysis);

% Decision arrows
\draw [arrow] (check) -- node[above] {Fail} (reprocess);
\draw [arrow] (reprocess) -- (bias);
\draw [arrow] (check) -- node[right] {Pass} (science);
\draw [arrow] (science) -- (analysis);

\end{tikzpicture}}
\caption{Flow chart for data processing pipline.}
\label{flowchart}
\end{figure}

The ``.raw-format" lunar camera image will be sent to the ground station at DSEL via the relay satellite. 
The raw image data will then undergo processing at the ground station and will be further processed 
using bias, dark, and flat-field images according to Equation~\ref{eq:1}. The modified data will be further 
checked for the astronomical World Co-ordinate System (WCS) calibration, sky background, and flux calibration. 
Figure~\ref{flowchart} shows the flowchart of the data processing pipeline.

\begin{equation} \label{eq:1}
Image_{\text{reduced}} = \frac{Image_{\text{raw}} - Bias_{\text{master}} - Dark_{\text{master}}}{Flat_{\text{master}}}
\end{equation}

\section{Simulated Performance \& Data Analysis}

\begin{figure}[ht]
\centering
\includegraphics[width=0.85\textwidth]{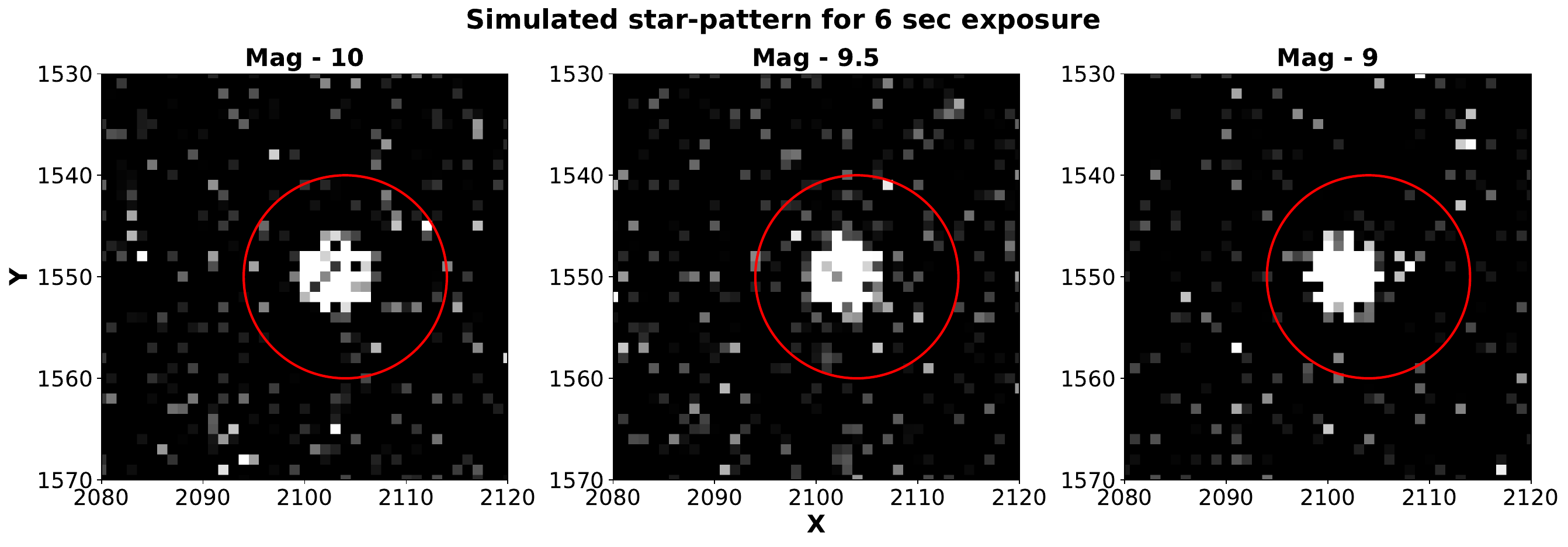}
\includegraphics[width=0.85\textwidth]{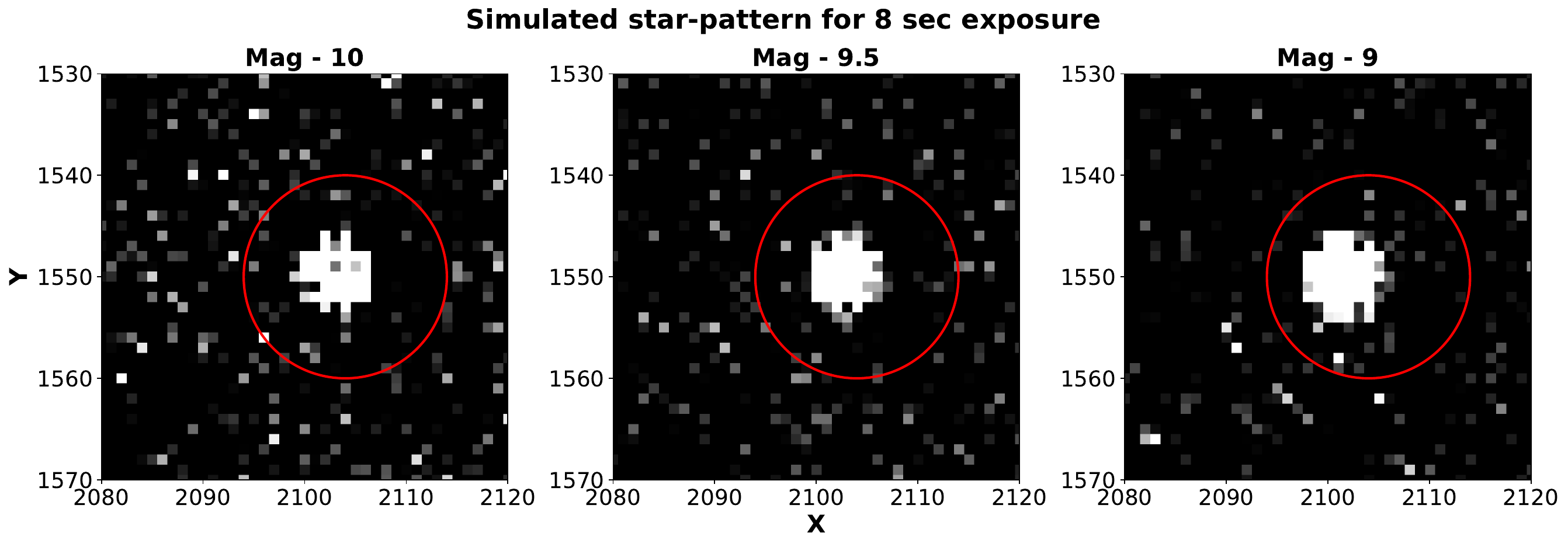}
\caption{Simulated stellar models and their variation with exposure and stellar magnitude. The red circle represents 
the 10-pixel-radius aperture used for aperture photometry.}
\label{star_simu}
\end{figure}

BISME has simulated stellar model images in .raw Bayer format, with dimensions of $4096\times3000$ and 16-bit depth 
(where the lower 12 bits contain the valid data). During the simulation, the variation of exposure and the magnitude 
of the star are considered. These simulated stellar models are treated as real observations. The samples of simulated 
data for 10, 9, and 9.5 magnitude stars are depicted in Fig.~\ref{star_simu}. 

We analysed these simulated ILO-C exposures and performed standard aperture photometry 
utilizing the Python-based image-processing package \texttt{Astropy Package Photutils} at a circular aperture of 10-pixel radius\cite{2013A&A...558A..33A, photutils} (\url{https://photutils.readthedocs.io/en/stable/user\_guide/aperture.html}). For each simulated  image, the $3-\sigma$ clipped background and its standard deviation are calculated. 
The background is subtracted from the image data during our analysis.

We performed the analysis for a 9-magnitude star at 6~sec and 8~sec exposure times (Fig.~\ref{star_simu}). 
During the 6~sec exposure time, the maximum intensity of 3573.24 over the background of $51.75\pm38.40$ is reported. 
In the case of 8~sec exposure time, the maximum intensity of 4897.83, over a background of $76.16\pm43.55$ is calculated. 
The corresponding values calculated for a 9.5 magnitude star (Fig.~\ref{star_simu}) are 2383.02 over the background of $84.98\pm48.34$ (6~sec exposure time), 3155.65, over a background of $92.35\pm50.72$ (8~sec), and 4004.75, over a background of $119.25\pm65.27$ (10~sec).
 For a 10-magnitude star, the exposure times of 4~sec, 6~sec, 8~sec, and 10~sec are considered. The corresponding values are 1400.45 over the background of $112.55\pm63.62$, 1566.31, over a background of $87.69\pm49.41$, 2309.34, over a background of $102.66\pm59.07$, and 2890.01, over a background of $98.99\pm59.62$, respectively.
Figure~\ref{star_simu} shows the background-subtracted simulated stellar models for stellar mag 10, 9.5, and 9, respectively, for the simulated exposure of 6 and 8 sec. These plots show the approximately linear correlation of exposure time and magnitude on the simulated intensity recorded by the lunar lander camera.

\begin{figure}[ht]
\centering
\includegraphics[width=0.65\textwidth]{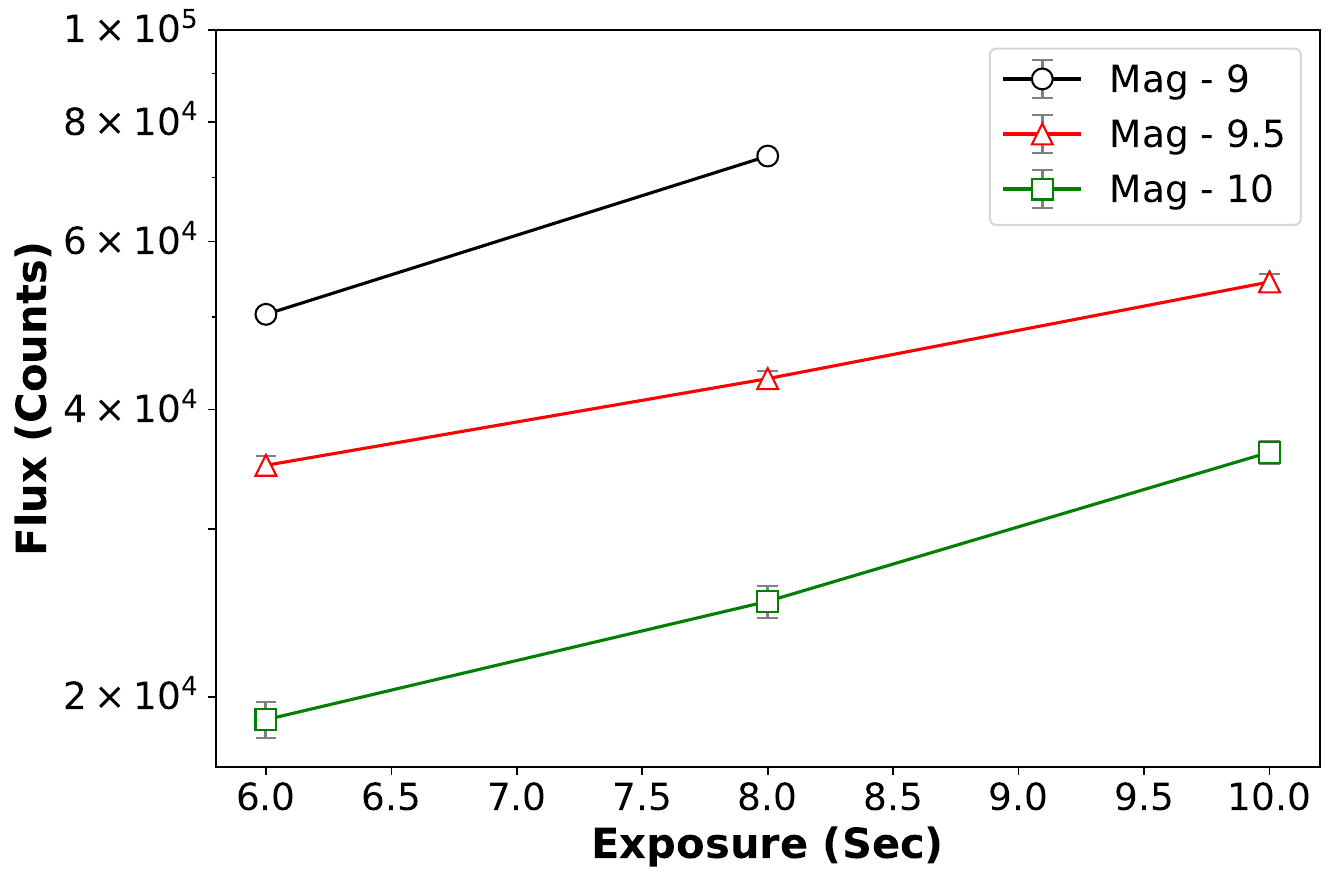}
\caption{Aperture photometry of the simulated stellar model.}
\label{aper}
\end{figure}

The background-subtracted photon fluxes are calculated from aperture photometry within a 10-pixel radius aperture 
for all the simulated stars. The 10 magnitude stellar model shows a variation in flux from $(18.93\pm0.82)\times10^3$, $(25.19\pm0.97)\times10^3$, and $(36.09\pm0.97)\times10^3$ during 6~sec, 8~sec, and 10~sec exposures, respectively. 
The 9.5 magnitude stellar model shows a variation in flux from
$(34.97\pm0.80)\times10^3$, $(43.10\pm0.84)\times10^3$, $(54.43\pm1.08)\times10^3$ during 6~sec, 8~sec and 10~sec exposures respectively. The 9 magnitude stellar model shows a variation in flux from
$(50.34\pm0.60)\times10^3$ to $(73.75\pm0.72)\times10^3$ 
during 6~sec and 8~sec exposures respectively. The flux comparison is shown in Fig.~\ref{aper}.

\begin{figure}[htbp]
\centering
\includegraphics[width=0.65\textwidth]{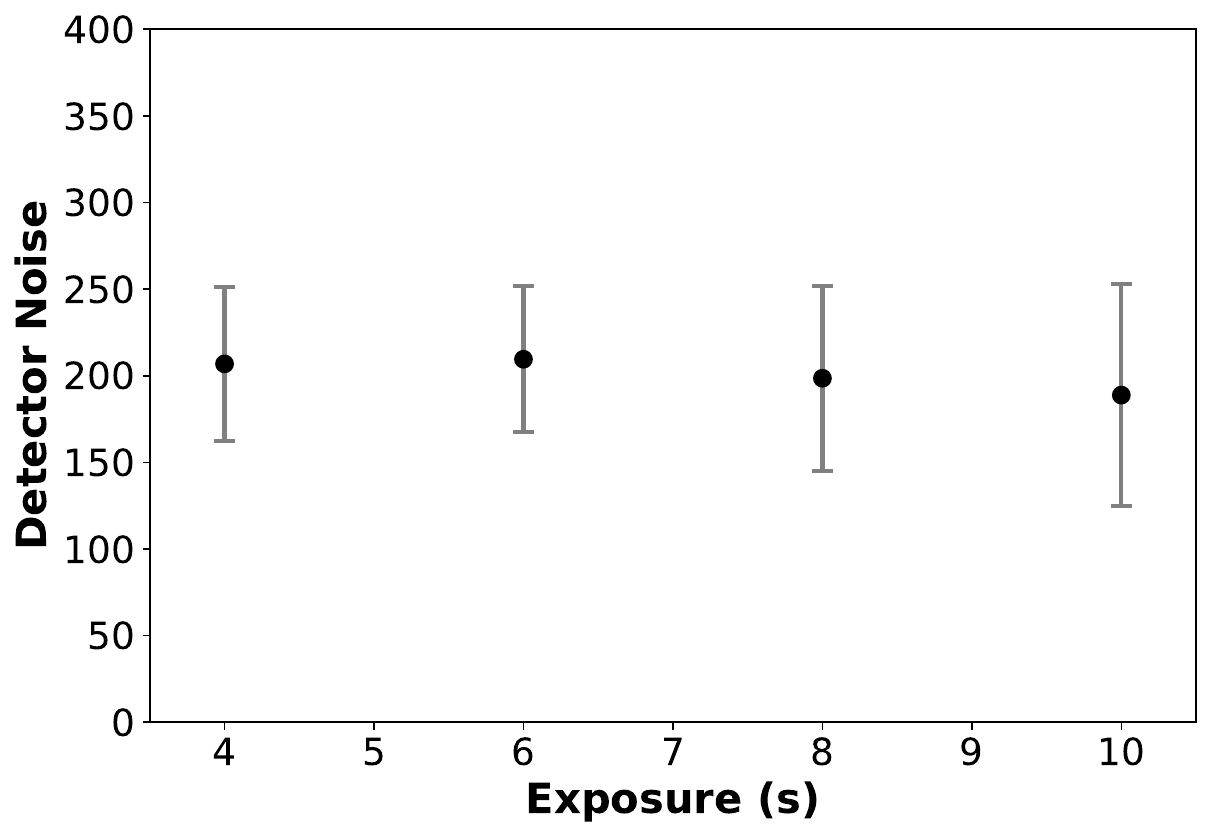}
\caption{Simulated noise for various exposures.}
\label{noise}
\end{figure}

The noise level of the detector readout system has also been measured. For exposures of 4~sec and 6~sec, 
the noise level is $207\pm44$ and $209\pm42$, respectively; it then decreases to $198\pm53$ and $189\pm64$ 
for the exposures of 8~sec and 10~sec, respectively. In Fig.~\ref{noise}, the detector noise shows a gradual 
reduction with increasing exposure time.

\section{Astronomical Capabilities}
A lunar camera of this character is limited in its astronomical capabilities by several 
fundamental instrumental properties. These are: 
\begin{enumerate}[label=(\roman*)]
\item The limited entrance aperture that  determines the light gathering power and hence potential 
astronomical depth (achieved in magnitudes) for point sources; 
\item  The camera's resolving power and hence its inherent angular resolution provided by the 
optics not limited by any atmosphere, i.e., how closely two equally bright stars can be to each other in 
arcseconds before we cannot determine there are two stars present; 
\item The relation to detector pixel size across the available field of field of view for actual 
arcseconds/pixel; 
\item The actual sensitivity and hence magnitude limit of the camera factoring in the overall 
efficiency of the optics and detector's quantum efficiency. In an ideal system the inherent 
angular resolution would be sampled by the detector pixels at the Nyquist rate.
\end{enumerate}

The astronomical science that can then be delivered depends on several further factors:
\begin{enumerate}[label=(\roman*)]
\item Total exposures that can be taken (over the camera lifetime, assumed currently as 7 earth 
days); 
\item The optimal exposure time determined to avoid star trails across pixels as the camera 
is fixed and to maximise depth; 
\item The total sky coverage  achieved and what part 
of the sky is actually observed; 
\item The panchromatic nature of the camera from 0.4 to 0.7~nanometres.
\end{enumerate}

An important additional limitation is the actual positioning achieved by the Chang'e 7 lander 
on the lunar surface. The orientation of the camera in azimuth and altitude on the lander, 
compared to nominal horizontal, will be determined by the actual landing site and how the lander 
settles onto the lunar surface. Here, regolith consistency will affect how the lander may sink 
asymmetrically into it. Additional factors that may affect camera performance are scattered light
(despite the camera baffling) and regolith dust that may get stirred up and fall onto 
the camera during the landing.

Consequently whether the camera ends being able to observe the Galactic Plane depends critically on 
how the lander will be actually situated. The desire to capture the Galactic plane rising up above the 
rim of the Shackleton crater was one of the prime drivers for selecting the widest possible field of view for 
the camera, so increasing the likelihood that this will be achieved.

With these points in mind we believe the following science capabilities are in principle possible:
i) Colour maps of the Milky way and, given the southern sky viewing available from the lunar 
South Pole region, hopefully the Galactic centre, and potentially the Large Magellanic Cloud. 
These will provide significant public outreach opportunities;
ii) Temporal  studies of certain kinds of bright, detectable variables whose periods might be 
well sampled during the mission lifetime and over the areas covered (results may be compatible with 
Chinese Antarctic Dome-A CSTAR program, for example); iii) Sensitivity to any novae in our own Galaxy or 
even SuperNovae explosions in any external galaxies that might, by chance, be imaged by the camera over its active lifetime;
iv) Sensitivity to near lunar objects (small asteroids) that may move across parts of the FoV.

\section{Discussion}

\begin{figure}[ht]
\centering
\includegraphics[width=0.75\textwidth]{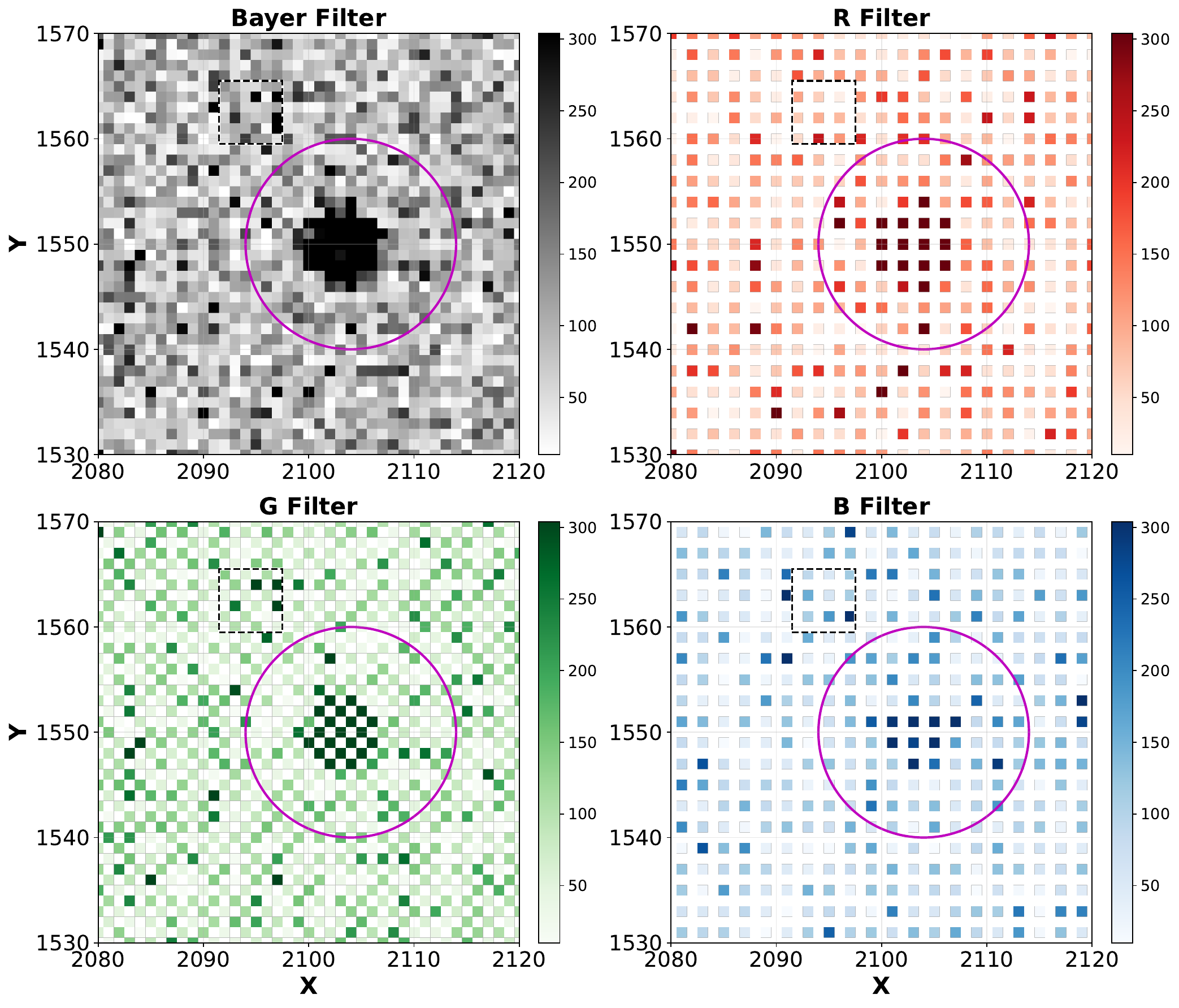}
\caption{Filter-dependent aperture photometry of the simulated 10 magnitude 10~sec exposure stellar model. The Bayer filter image is shown in reverse gray scale. The magenta circle represents the 10-pixel-radius aperture used for photometric analysis. The black square is shown for reference to explain the RGGB pattern.}
\label{color}
\end{figure}

Here we report a further possibility for in-depth scientific analysis of the simulated stellar images. The sCMOS sensor stores the image data in \texttt{RGGB Bayer} filter patterns. In Fig.~\ref{resp}, the spectral response function of the individual R, G, and B pixels is shown. In Fig.~\ref{color}, the mosaic image of a simulated stellar model from Bayer filter and individual R, G, B filters is shown. The aperture photometry is performed on each mosaic image using a 10-pixel radius aperture. In Table~\ref {mos_phot}, the comparison of background-subtracted photon fluxes within the selected aperture is given from mosaic images of individual filters. G-filter flux is divided by 2 in order to normalize with R and B filters. Similar calculation is repeated for 9.5- and 10-magnitude stellar models over a 10-second exposure.  

\begin{table}[ht]
\caption{Aperture photometry of filter-dependent mosaic image}
\label{mos_phot}
\begin{center}
%\scriptsize 
\begin{tabular}{|c|c|c|c|c|c|}
\hline\hline
\textbf{Magnitude} & \textbf{Exposure} & \textbf{Bayer} & \textbf{R} & \textbf{G} & \textbf{B} \\
\textbf{} & \textbf{(Sec)} & \textbf{($\times10^3~counts$)} & \textbf{($\times10^3~counts$)} & \textbf{($\times10^3~counts$)} & \textbf{($\times10^3~counts$)} \\\hline
10 & 10 & $36.09\pm2.06$ & $10.13\pm0.05$ & $11.07\pm0.69$ & $3.81\pm1.11$ \\
9.5 & 10 & $54.43\pm2.43$ & $15.00\pm1.22$ & $17.87\pm0.82$ & $3.72\pm1.31$ \\
\hline
\end{tabular}
\end{center}
\end{table}

The scientific analysis of the lunar lander camera will add extra scientific significance to optical astronomy from the lunar surface. It will enable us to observe the lunar optical sky across multiple optical filter bands. A similar multi-filter optical-wavelength full sky astronomical survey is also conducted by the very recent \texttt{SPHEREx} Satellite Mission of NASA \cite{2026ApJ...999..139B}.

\section{Conclusion}
In this paper, we have reported the scope of a new pathfinder optical astronomy observation strategy from the lunar surface. The expected image quality and the first version of a photometry pipeline for analyzing captured data are introduced. Given the noise and dark-current responses of the camera-telescope to variations in temperature and exposure time, the instrument is expected to record valuable data during the mission. It should be emphasized that this kind of observation strategy will not only provide a safer and cost-effective test bed for new technological instruments on the lunar surface, but also enable the very latest optical astronomy observations. The lunar surface-based astronomical facility will definitely prove to be a safer testing ground than space missions, as it will offer greater opportunity for future upgrades to instrument hardware. It is much safer and cost-effective than other space missions.

\subsection*{Disclosures}
The authors declare that there are no financial interests, commercial affiliations, or other potential conflicts of interest that could have influenced the objectivity of this research or the writing of this paper

\subsection* {Code, Data, and Materials Availability} 
The calibration and simulation data are generated by BISME. These data are available to members of the ILO-C collaboration.  

\subsection* {Acknowledgments}
We acknowledge BISME and DSEL for their technical support. 
This project acknowledges the funding support from HKU-LSR and ILOA.

%%%%% References %%%%%

\bibliography{reference}   % bibliography data in report.bib
\bibliographystyle{spiejour}   % makes bibtex use spiejour.bst

\end{document}